\documentclass[12pt]{article}

\usepackage{amsmath}
\usepackage{graphicx,psfrag,epsf}
\usepackage{enumerate}
\usepackage{natbib}

\newcommand{\blind}{0}

\usepackage[utf8]{inputenc}
\usepackage{amssymb}
\usepackage{float,color}
\usepackage{graphicx}
\usepackage{stackrel}
\usepackage{comment}
\usepackage{amsmath} 
\usepackage{amsmath}
\usepackage{bm}
\usepackage{amssymb,amsmath,amsthm,bm} 
\usepackage{bbm}
\usepackage{subfig}
\usepackage{tabularx}
\usepackage{amsmath}

\usepackage[margin=1in]{geometry}
\usepackage{setspace}
\usepackage{booktabs} 
\usepackage{threeparttable,array,booktabs}
\usepackage{lipsum}

\usepackage{algorithm,algcompatible,amsmath}
\algnewcommand\INPUT{\item[\textbf{Input:}]}%
\algnewcommand\OUTPUT{\item[\textbf{Output:}]}%

\usepackage{natbib}
\setcitestyle{authoryear,open={(},close={)}}

\newtheorem{theorem}{Theorem}[section]

\newtheorem{prop}{Proposition}

\title{Analysis of professional basketball field goal attempts via a Bayesian matrix clustering approach}

\author{Fan Yin$^{1}$, Guanyu Hu$^{2}$, Weining Shen$^{1}$   \\
        \small $^{1}$University of California - Irvine, Irvine, CA, 92697 \\
        \small $^{2}$University of Missouri - Columbia, Columbia, MO, 65211 \\
}
 
\date{}

\begin{document}

\def\spacingset#1{\renewcommand{\baselinestretch}%
{#1}\small\normalsize} \spacingset{1}

\if0\blind
{
  \title{\bf Analysis of professional basketball field goal attempts via a Bayesian matrix clustering approach}
  \author{Fan Yin\\
    Department of Statistics, University of California - Irvine\\
    Guanyu Hu \\
    Department of Statistics, University of  Missouri - Columbia \\
    and \\
    Weining Shen\thanks{
    Corresponding author, Email: weinings@uci.edu}\hspace{.2cm} \\
    Department of Statistics, University of California - Irvine}
  \maketitle
} \fi

\if1\blind
{
  \bigskip
  \bigskip
  \bigskip
  \begin{center}
    {\LARGE\bf Title}
\end{center}
  \medskip
} \fi

\bigskip
\begin{abstract}
We propose a Bayesian nonparametric matrix clustering approach to analyze the latent heterogeneity structure in the shot selection data collected from professional basketball players in the National Basketball Association (NBA). The proposed method adopts a mixture of finite mixtures framework and fully utilizes the spatial information via a mixture of matrix normal distribution representation. We propose an efficient Markov chain Monte Carlo algorithm for posterior sampling that allows simultaneous inference on both the number of clusters and the cluster configurations. We also establish large-sample convergence properties for the posterior distribution. The excellent empirical performance of the proposed method is demonstrated via simulation studies and an application to shot chart data from selected players in the NBA’s 2017–2018 regular season.
\end{abstract}

\noindent%
\spacingset{1.45}
\section{Introduction}\label{sec:intro}
Sports analytics have received an increasing interest in statistics community and they continue to offer new challenges as a result of ever-increasing data sources. Conventional statistical research for sports analytics was mainly concerned with forecasting the game results, such as predicting the number of goals scored in soccer matches \citep{dixon1997modelling, karlis2003analysis,baio2010bayesian}, and the basketball game outcomes \citep{carlin1996improved,caudill2003predicting,cattelan2013dynamic}. More recently, fast development in player tracking technologies has greatly facilitated the data collection \citep{albert2017handbook}, and in turn substantially expanded the role of statistics in sports analytics, including granular evaluation of player/team performance \citep{cervone2014pointwise,franks2015characterizing, cervone2016multiresolution,wu2018modeling}, and in-game strategy evaluation \citep{fernandez2018wide,sandholtz2019measuring}. 

In professional basketball research, shooting pattern remains to be a fundamental metric for evaluating players' performance and has aroused great interest among statisticians. Shot charts, as graphical representations of players' shot locations, provide an excellent tool to summarize and visualize shooting patterns for players. To account for the spatial correlation in shot chart data, several spatial statistical models have been studied in the literature. For example,  \citet{reich2006spatial} proposed a multinomial logit regression model with spatially varying coefficients to quantify the effects of several hand-crafted features (e.g., home or away games, average number of blocks made by the defensive player, the presence of certain other teammates) on the probability of making a shot over different basketball court regions. More recently, spatial point process \citep{miller2014factorized, jiao2019bayesian} has emerged as a promising direction for shot chart data analysis in recognition of the randomness nature of shot locations. In those works, it is common to first summarize the shot charts as intensity matrices of the underlying point process and then conduct a regression analysis of pre-specified artificial baseline shooting patterns on game outcomes. A common finding in these studies is that the shooting behaviors are {\it highly heterogeneous} among different players, which calls for a clustering analysis towards a deeper understanding of the player-level heterogeneity and the improvement of existing statistical models by incorporating the latent clustering structure. To date, most existing clustering approaches for shot chart data analysis are distance-based (e.g., $K$-means and hierarchical clustering), and hence lacking a probabilistic interpretation. A model-based clustering approach was proposed in \citet{hu2020bayesian} based on calculating the similarity matrix between intensity matrices of players' shot charts. However, that method still lacks an intuitive model interpretation for the clustering results since the clustering is performed based on the similarity matrix, rather than the intensity matrices.  

The main goal of this paper is to fill this gap by introducing a novel Bayesian model-based clustering approach for learning the basketball players' heterogeneity via their shot chart data analysis. The key novelty of our method starts from treating each shot chart as a {\it matrix}, i.e., the basketball court is divided into a few rectangle regions and the number of shots (or the intensity of the underlying spatial point process) over those regions are represented as elements in the corresponding matrix. The immediate benefit of treating each sampling unit (shot chart) as a matrix is that it automatically takes account for the spatial structure information in the analysis. Moreover, it allows us to conveniently extend the classical Gaussian mixture model (for vectors) for clustering matrix-valued shot chart data purpose. Gaussian mixture models (and mixture models in general) have been widely used in many applications thanks to their convenient probabilistic interpretation and elegant computational solutions such as the expectation maximization (EM). However, mixture models for matrix-valued data have received little attention until recently.  
Most existing works \citep{viroli2011finite,thompson2020classification,gao2018regularized} are based on the EM framework, which requires pre-specifying the number of clusters while the inference cannot be easily conducted for clustering outputs over different cluster numbers. A Bayesian approach was proposed in \citet{viroli2011model}  by imposing a prior on the number of clusters and drawing posterior samples with a birth and death Markov chain Monte-Carlo algorithm \citep[BDMCMC;][]{stephens2000bayesian}. However, that approach requires a careful parameter tuning process in BDMCMC and the computation does not scale up with the size of matrices. To date, it remains challenging to conduct efficient Bayesian inference for matrix-valued mixture models due to the large parameter space (e.g., the number of parameters is at least of order $O(p^2 + q^2)$ 
for $p \times q$ matrices) and the fact that the parameter space is not fixed as the number of clusters varies. Moreover, there is a lack of understanding of the theoretical properties for these mixture models. 

Our methodology development is directly motivated by solving the aforementioned challenges. In particular, we propose MFM-MxN, which is a novel nonparametric Bayesian mixture model of matrix normal distributions (MxN) under the mixture of finite mixtures framework \citep[MFM;][]{miller2018mixture}. The main idea is to represent each cluster (of shot charts) by a matrix normal distribution and allow the number of clusters to be random. We develop a Gibbs sampler that enables efficient full Bayesian inference on the number of clusters, mixture probabilities as well as other modeling parameters. We demonstrate its excellent numerical performance through simulations and an analysis of the NBA shot chart data. In addition, we establish a consistency result for the posterior estimates of the cluster number and the associated modeling parameters.

Our proposed method is unique in the following aspects. First, the idea of representing each player's shot chart as an intensity matrix and formally introducing the concept of {\it matrix data analysis} for solving clustering problem is novel. In fact this idea and our proposed approach are widely applicable to general sports applications such as baseball and football studies, and provide a valuable alternative to the existing literature that mainly relies on spatial methods. Secondly, by adopting a full Bayesian framework, the clustering results yield useful probabilistic interpretation. Moreover, the developed posterior sampling scheme also renders efficient computation and convenient inference compared to all the other methods for modeling matrix-valued data in the literature. Thirdly, our theoretical result is among the first of its kind for mixture models of matrix-variate distributions. The posterior consistency result not only provides a theoretical justification for the excellent empirical performance (e.g., high probability of selecting the correct number of clusters), but also connects to the existing theoretical findings on mixture models in general (for vector-valued data).

\section{Motivating Data Example}\label{sec:motivating_data}
\begin{figure}[htp]
    \centering
    \includegraphics[width=0.8\textwidth]{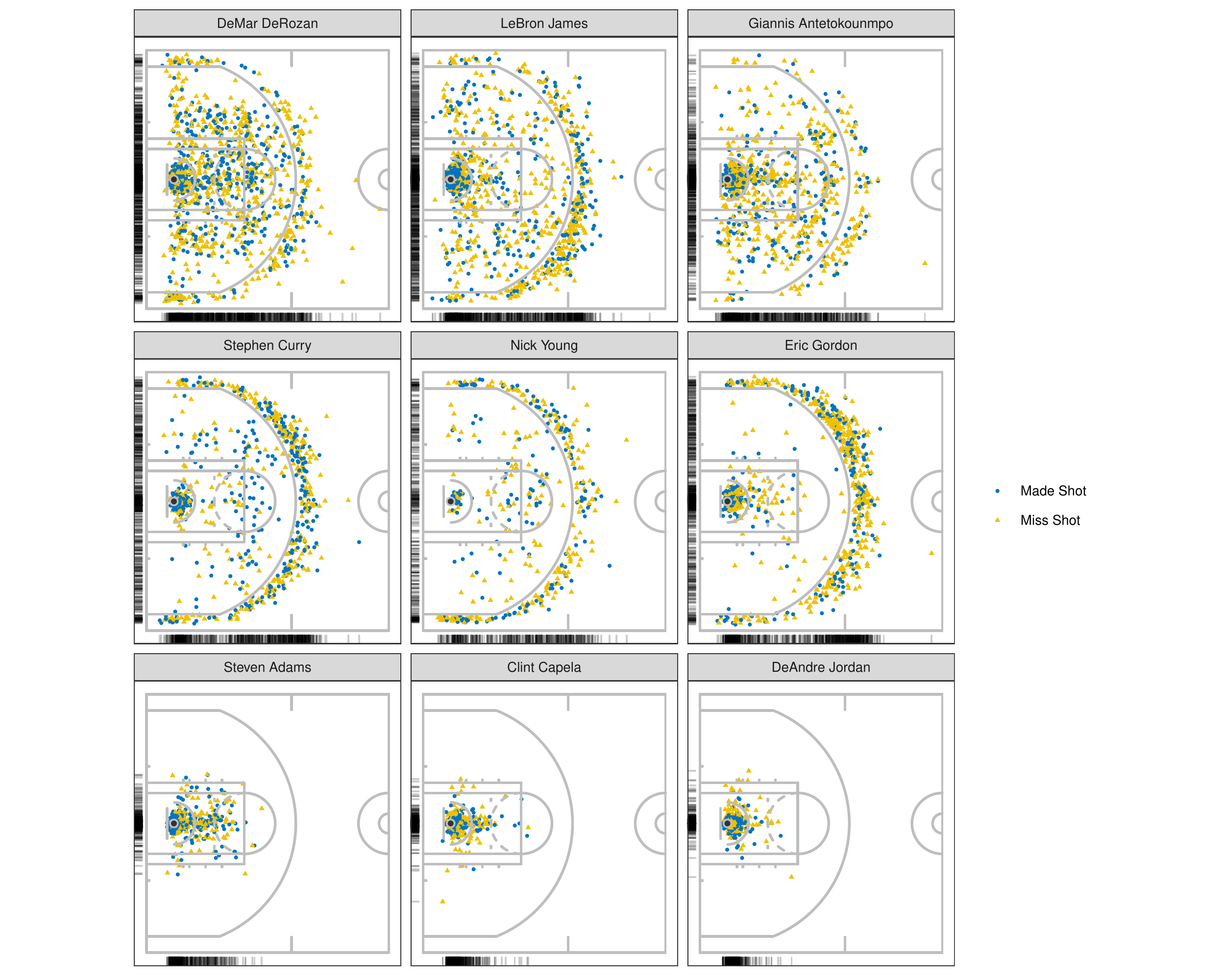}
    \caption{Shot charts for selected NBA players}
    \label{fig:data_presnet}
\end{figure}

We consider a dataset consisting of locations of field goal attempts (FTA) from the offensive half court in all 82 games during 2017–2018 National Basketball Association (NBA) regular season. Following \citet{hu2020bayesian}, we focus on 191 players who have made more than 400 FTAs in that season. The rookie year players, such as Lonzo Ball and Jayson Tatum, are not included in our analysis. All the shooting locations are in a 47 ft (baseline to mid court line) by 50 ft (sideline to sideline) rectangle, which is the standard court size for NBA games. We select nine players (DeMar DeRozan,  LeBron James, Giannis Antetokounmpo, Stephen Curry, Nick Young, Eric Gordon, Steven Adams, Clint Capela, DeAndre Jordan) and visualize their shot charts in Figure~\ref{fig:data_presnet}. From this figure, we observe a clear heterogeneity pattern, e.g., the  three players in the first row have a more balanced spatial location pattern in their FTAs than those from the other six players; the three players in second row have more FTAs around 3-pt line; and the three players in last row have more FTAs near the basket. Those observations seem closely related to their positions and playing styles in the game. Our goal in this paper is to synthesize these empirical findings through a formal model-based clustering approach. 

\section{Method}\label{sec:method}
In this section, we first give a brief review of log Gaussian Cox process and matrix normal distribution, and then present our Bayesian matrix normal mixture model in Section \ref{sec:MxN_MFM}. 
\subsection{Log Gaussian Cox Process}\label{sec:lgcp}
Consider a collection of 2D spatial locations $\bm{S} = \{\bm{s}_1, \bm{s}_2, \dots, \bm{s}_N\}$ over a study region
$\mathcal{B} \subset \mathbb{R}^2$.
It is common to represent the underlying spatial pattern by a spatial point process characterized by a quantity called intensity. Formally, within a region $\mathcal{B}$, the intensity at location $\bm{s}\in \mathcal{B}$ is defined as
\begin{equation*}
\lambda(\bm{s}) = \lim_{|d  \bm{s}\rightarrow
    0|}\left(\frac{\textrm{E}[N(d\bm{s})]}{|d  \bm{s}|} \right),
\end{equation*}
where~$d\bm{s}$ is an infinitesimal region around~$\bm{s}$, $|d \bm{s}|$
represents its area, and $N(d \bm{s})$ denotes the number of events that
happens over $d\bm{s}$. A spatial Poisson point process is a process such that the number of events/points in any subregion $A\subset \mathcal{B}$ follows a Poisson
distribution with mean
$\lambda(A) = \int_{A}\lambda(\bm{s})d \bm{s}$
for some function $\lambda(\cdot)$. Similarly with the Poisson distribution, a Poisson process $\mathcal{PP}(\lambda(\cdot))$ satisfies $\text{E}(N_{\bm{S}}(A))=\text{Var}(N_{\bm{S}}(A))=\lambda(A)$. A homogeneous
Poisson process (HPP) assumes
$\lambda(\bm{s})=\lambda$, i.e., the intensity is a constant over the entire region 
$\mathcal{B}$. A more realistic case is to let $\lambda(\bm{s})$ vary spatially, which leads to a nonhomogeneous Poisson process. 

\begin{figure}[htp]
    \centering
    \includegraphics[width=0.75\textwidth]{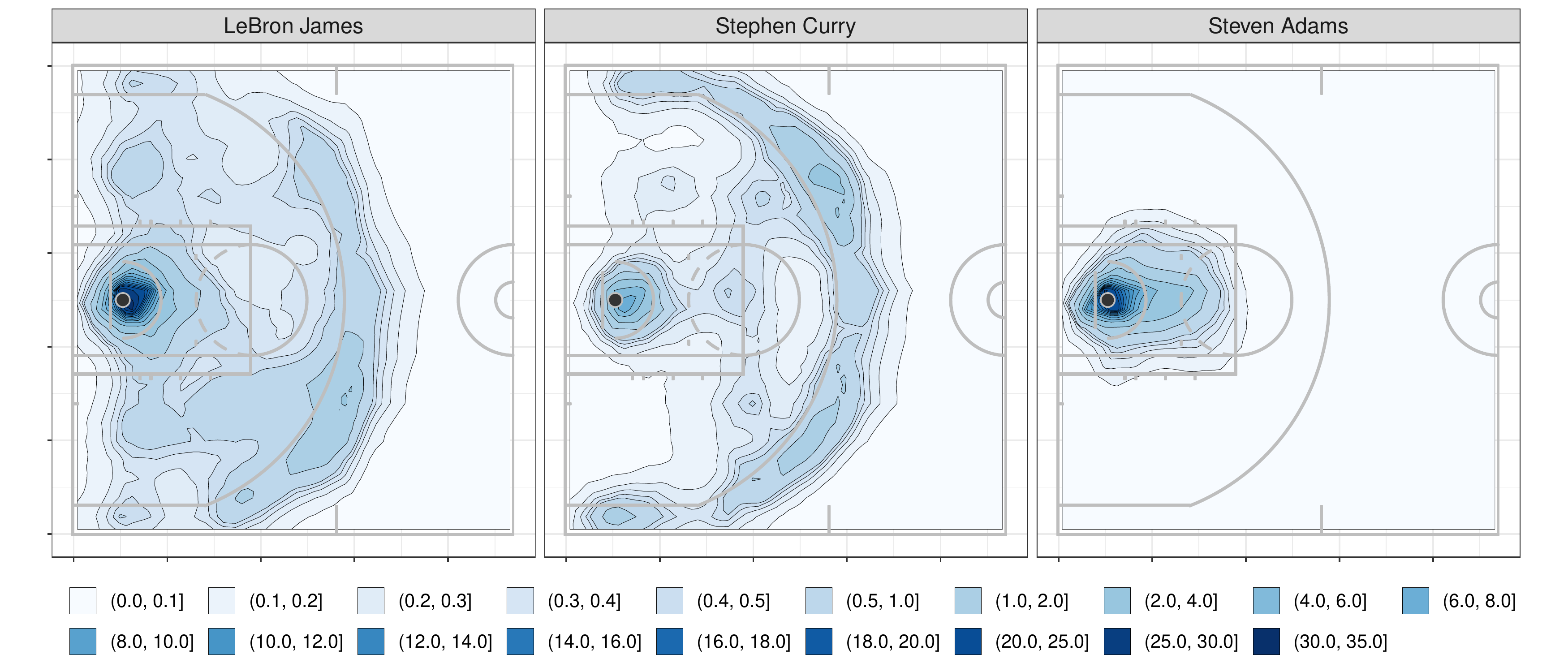}
    \caption{Estimated Intensity Maps for three selected players}
    \label{fig:estimated_intensity}
\end{figure}

Among the class of Poisson processes, log Gaussian Cox process (LGCP) has received a lot of attention in practice thanks to its flexibility and easy interpretability. A LGCP is a doubly-stochastic Poisson process with a correlated and spatially-varying intensity \citep{moller1998log}, defined as follows, 
\begin{equation}
\label{eq:h_lgcp}
	\bm{S}  \sim \mathcal{PP}(\lambda(\cdot)), ~~
	\lambda(\cdot)  =\exp(Z(\cdot)), ~~
	Z(\cdot)  \sim \mathcal{GP}(0,k(\cdot,\cdot)),
\end{equation} 
where $Z(\cdot)$ is a zero-mean Gaussian process with covariance kernel $k(\cdot,\cdot)$. From \eqref{eq:h_lgcp}, the LGCP can be viewed as an exponentiated Gaussian process, such as a Gaussian random field \citep[GRF;][]{rasmussen2003gaussian}, which assumes that the log intensities at different spatial
locations are normally distributed, and spatially correlated. To relate to our basketball shot chart data discussed in Section \ref{sec:motivating_data}, for all the players of interest, we can model their shot charts $\bm{S}^{(1)},\bm{S}^{(2)},\ldots,\bm{S}^{(n)}$ through a LGCP and estimate their associated intensity functions, denoted by $\lambda^{(1)}(\cdot),\lambda^{(2)}(\cdot),\ldots,\lambda^{(n)}(\cdot)$. This step can be conveniently implemented using integrated nested Laplace approximation \citep[INLA;][]{rue2009approximate}. See more details of implementation in \citet{cervone2016multiresolution} and \citet{hu2020bayesian}. For illustration, we plot the estimated intensity maps for three selected players in Figure~\ref{fig:estimated_intensity}. 


\subsection{Matrix normal distribution}\label{sec:matrix_data}
Next we provide a brief review of matrix normal distribution. Consider a $p \times q$ random matrix $Y$. We say $Y$ follows a matrix-variate normal distribution (MxN) with parameters $M$, $U$ and $V$, denoted by, $Y \sim \mathcal{N}_{p,q}(M, U, V)$, if it has the following probability density function
\begin{equation}
\label{eq:MxN}
f(Y; M, U, V) = \frac{\exp( -\frac{1}{2} \text{tr}[V^{-1} (Y - M)^{\intercal} U^{-1} (Y - M)] )}{ (2\pi)^{pq/2} |V|^{p/2} |U|^{q/2} },
\end{equation}
where matrix $M \in \mathcal{R}^{p \times q}$ is the mean of $Y$, and $|\cdot|$ denotes the matrix determinant. Here positive definite matrices $U \in \mathcal{R}^{p \times p}$ and $V \in \mathcal{R}^{q \times q}$ are row-wise covariance and column-wise covariance parameters, describing the covariances between, respectively, each of the $p$ rows and the $q$ columns of $Y$. It is clear from \eqref{eq:MxN} that the matrix normal distribution can be viewed as a multivariate normal distribution with a Kronecker product covariance structure \citep{gupta1999matrix}, that is, $Y \sim \mathcal{N}_{p,q}(M, U, V)$ is equivalent to $\text{vec}(Y) \sim \mathcal{N}_{pq}(\text{vec}(M), V \otimes U)$, where $\text{vec}(\cdot)$ is a vectorization operator that stacks all the columns in a matrix into a tall column vector. Since $V \otimes U = (\frac{1}{a} V) \otimes (a U)$ for any $a \neq 0$, we impose a constraint $\text{tr}(V) = q$ for  model identifiability purpose. 

From the definition of the matrix normal distribution, we can see that it enjoys a parsimonious covariance structure. By representing a $(pq) \times (pq)$ covariance as the Kronecker product of a $p \times p$ and a $ q \times q$ covariance matrix, it effectively reduces the number of unknown parameters from $pq (pq +1)/2$ to $\{p(p+1) + q(q+1) \}/2$. Moreover, it provides a useful interpretation  by projecting the spatial variability onto column and row directions, which can be viewed as a spatial version of the analysis of variance (ANOVA) model. For the basketball shot chart data, it is natural to divide the offensive half court equally into rectangle regions and represent the measurements (e.g., number of shots being made by a player) over those regions in a matrix form. Moreover, we can model the logarithm of the corresponding intensity function over the matrix by a matrix normal distribution. It is also worthy mentioning that there are other useful distributions defined for matrix-valued data, such as matrix-variate t distribution \citep{thompson2020classification}. Our proposed Bayesian mixture model of matrix normal distributions can be naturally extended to those distributions; and we  focus on matrix normal distribution here for its convenience and easy interpretation. 

\subsection{Bayesian matrix normal mixture model}\label{sec:MxN_MFM}
To account for the potential heterogeneity in the matrix-valued data, we propose a Bayesian mixture model where each mixture component is represented by a matrix normal distribution. More specifically, suppose that there are a total number of $K$ clusters, with weights $\pi_1,\ldots,\pi_K$, and each mixture follows a different matrix normal distribution. Then we adopt the mixture of finite mixtures (MFM) framework \citep{miller2018mixture} by assigning prior distributions on those unknown model parameters as follows, 
\begin{align} 
\label{eq:MFM_MxN} 
& K \sim p_{K}, \ \ p_{K} \ \ \text{is a p.m.f on} \ \mathbb{N}^{+} = \left\{1,2,\ldots \right\}, \nonumber \\
& \pi = (\pi_1,\ldots,\pi_k) \sim \text{Dir}_{k}(\gamma,\ldots,\gamma), \ \ \text{given} \ K = k, \nonumber \\
& P(Z_i = j) = \pi_j \ \ \text{for every} \ \ i=1,\ldots,n, ~\text{and}~ j=1,\ldots,k, \ \ \text{given} \ \pi, \nonumber \\
& M_1,\ldots, M_k \stackrel{i.i.d}{\sim}  \mathcal{N}_{p,q}(M_{0} , \Sigma_{0}, \Omega_{0}) \ \ \text{given} \ K = k, \nonumber  \\
& U_{1},\ldots,U_{k} \stackrel{i.i.d}{\sim} \mathcal{IW}_{p}(2 \alpha, (2\beta)^{-1} ) \ \ \text{given} \ K = k, \nonumber \\ 
& V_{1},\ldots,V_{k} \stackrel{i.i.d}{\sim}  \mathcal{IW}_{q}(2\psi, (2\rho)^{-1} ) \ \ \text{given} \ K = k, \nonumber \\ 
& Y_{i} \sim \mathcal{N}_{p,q}(M_{Z_{i}}, U_{Z_{i}}, V_{Z_{i}}) \ \text{independently for} \ i=1,\ldots,n, \ \text{given} \ \bm{\Theta} \ \text{and} \ Z_{1}, \ldots, Z_{n}, 
\end{align} 

where $Z_1,\ldots,Z_n$ are cluster membership indicators that take values in $\{1,\ldots,K\}$ for each observation $Y_i$, $\bm{\Theta} = (\bm{\Theta}_{1}, \ldots, \bm{\Theta}_{K})$ and $\bm{\Theta}_{k} = (M_{k}, U_{k}, V_{k}), k = 1,\ldots,K$ are the collection of the parameters in the matrix normal distribution, and $\gamma$, $\psi, \rho$ are hyper-parameters. Here $\mathcal{IW}_{p}(\nu, S^{-1})$ means an inverse-Wishart distribution on $p \times p$ positive definite matrices with degree of freedom $\nu  (\nu > p-1)$ and scale parameter $S$, the probability density of which is proportional to $|\Sigma|^{-(\nu+p+1)/2} \exp(-\text{tr}(S \Sigma^{-1}/2))$. In our data analysis, $Y_i$'s are the log intensities $\log(\hat{\lambda}^{(i)}(\cdot))$ of LGCPs obtained in Section \ref{sec:lgcp}. We follow the convention to choose $p_{K}$ as a Poisson($\tau=1$) distribution truncated to take only positive values \citep{miller2018mixture,geng2019bayesian}. The prior distributions for $\bm{\Theta}_{k}$'s are specified to facilitate Bayesian inference via Gibbs sampling by taking advantage of the Normal-Normal and Normal-inverse-Wishart conjugacy. We will discuss more details about the numerical implementation in the later sections. 

The matrix normal mixture model has been previously studied in \citet{viroli2011finite} under the EM framework and in \citet{gao2018regularized} by imposing regularization on the mean structure for sparsity structure. However, in both works, it remains challenging to conduct full inference on the number of clusters and the cluster parameters simultaneously. \citet{viroli2011model} considered a Bayesian matrix normal mixture model and proposed to use birth and death MCMC algorithm for posterior inference. However, that method does not scale up to the size of the matrix and the theoretical property of the Bayesian estimators remains largely unknown. We will provide more details about computation and theoretical results, and highlight our contributions in the next two sections. 



\section{Bayesian Inference}\label{sec:inference}
In this section, we present a Gibbs sampler that enables efficient Bayesian inference for the proposed model and adopt the Dahl's method \citep{dahl2006model} for post-processing the MCMC outputs.

\subsection{MCMC Algorithm}\label{sec:mcmc}
By exploiting the conditional conjugacy property in model specification \eqref{eq:MFM_MxN}, we derive a collapsed Gibbs sampler algorithm for efficient Bayesian inference. Detailed derivations of the full conditionals are provided in Sections S3 and S4 of the Supplementary Materials.

For the basketball application, we find it plausible to assume that different mixture components share the same covariance structure, that is, $U_{1} = \cdots = U_{K} = U$ and $V_{1} = \cdots = V_{K} = V$. Extension to allow distinct covariances for different clusters is possible by considering auxiliary parameters when updating indicator variables $Z_i, i=1,\ldots,n$ using the method in \citet{neal2000markov}. Based on the Algorithm 2 of \citet{neal2000markov}, we obtain the following proposition that provides the full conditional distribution of $Z_i, i=1,\ldots,n$ while collapsing the number of clusters $K$. 
\begin{prop}
\label{prop:Z_update}
The full conditional distributions $P(Z_i | Z_{-i}, \bm{\Theta} )$ is given by
\begin{equation}
 P(Z_i | Z_{-i}, \bm{\Theta} ) \propto \begin{cases} 
          (\#c + \gamma) f(Y_{i} | M_{Z_{i}}, U, V) & \text{at an existing cluster} \ $c$ \\
          \frac{V_{n}(\#\mathcal{C}_{-i} + 1)}{V_{n}(\#\mathcal{C}_{-i})} \gamma m(Y_{i} | U, V) & \text{if} \ c \ \text{is a new cluster}
      \end{cases} ,    
\end{equation}
where $\#c$ refers to the cardinality of the cluster labeled as $c$, $f(Y_{i} | M_{Z_{i}}, U, V)$ is the density function of MxN defined in \eqref{eq:MxN}, $V_{n}(t)$ is a coefficient for the partition distribution defined as
$$ V_{n}(t) = \sum_{k=1}^{\infty} \frac{k_{(t)}}{(\gamma k)^{(n)}} p_K(k), $$
with $k_{(t)} = k(k-1)\ldots(k-t+1)$, $(\gamma k)^{(n)} = \gamma k (\gamma k + 1)\ldots(\gamma k + n - 1)$, $\mathcal{C}_{-i}$ represents a partition of the set $\left\{1,2,\ldots,n\right\} \setminus \{i\}$, and let $\#\mathcal{C}_{-i}$ denote the number of blocks in the partition $\mathcal{C}_{-i}$. Also, we define $m(Y_{i} | U, V)$ as 
\begin{equation*}
      \frac{\exp(-\frac{1}{2}[\text{vec}(Y_{i})^{\intercal}(V^{-1} \otimes U^{-1})\text{vec}(Y_{i}) + \text{vec}(M_{0})^{\intercal} (\Omega_{0}^{-1} \otimes \Sigma_{0}^{-1}) \text{vec}(M_{0}) - \tilde{\mu}^{\intercal} \tilde{\Sigma}^{-1} \tilde{\mu} ])}{(2\pi)^{pq/2}|V|^{p/2} |U|^{q/2} |\Omega_{0}|^{p/2} |\Sigma_{0}|^{q/2}} |\tilde{\Sigma}|^{pq/2},
\end{equation*}

where $\tilde{\Sigma}^{-1} = V^{-1} \otimes U^{-1} + \Omega_{0}^{-1} \otimes \Sigma_{0}^{-1} $ and $\tilde{\mu} = \tilde{\Sigma} [ (V^{-1} \otimes U^{-1}) \text{vec}(Y_{i}) + (\Omega_{0}^{-1} \otimes \Sigma_{0}^{-1}) vec(M_{0}) ]$.


\end{prop}
The derivation of $m(Y_{i} | U, V)$ in Proposition \ref{prop:Z_update} is given in the Supplementary Materials. Our collapsed Gibbs sampler algorithm for proposed model is summarized as Algorithm 1 in Section S5 of the Supplementary Materials.


 
We make the following recommendations for hyperparameter values in priors:
\begin{itemize}
    \item $\alpha = (p+1)/2$, $\psi = (q+1)/2$, $(2\beta)^{-1} = I_{p}$, $(2\rho)^{-1} = I_{q}$, which ensure the prior distributions for covariance matrices to be fairly diffuse while scale parameters $2\beta$, $2\rho$ are chosen to possess the simplest possible forms.
    \item $\gamma = 3$, which puts low probability on small group sizes. 
    \item Set $M_0$ as the (element-wise) middle point of the observations.
    \item $\Sigma_0 = \text{diag}(\sigma_{1}^{2}, \ldots, \sigma_{p}^{2})$, $\Omega_0 = \text{diag}(\omega_{1}^2,\ldots,\omega_{q}^2)$, where $\sigma_{1},\ldots,\sigma_{p}$ and $\omega_{1},\ldots,\omega_{q}$ are equal to half of the ranges along the respective rows and columns.
\end{itemize}

Numerical experiments have confirmed that the above hyperparameters work well, and hence these values will be used for all simulation studies and case studies in this paper. 

\subsection{Post MCMC Inference}\label{sec:post_mcmc}
We carry out posterior inference on the group memberships using Dahl's method \citep{dahl2006model}, which proceeds as follows,

\begin{itemize}
    \item \emph{Step 1.} Define membership matrices $\mathcal{A}^{(l)} =(\mathcal{A}^{(l)}(i,j))_{i,j \in \left\{1,\ldots,n\right\} } =  (\mathbbm{1}(Z_{i}^{(l)} = Z_{j}^{(l)}))_{n \times n}$, where $l = 1, \ldots, L$ is the index for the retained MCMC draws after burn-in, and $\mathbbm{1}(\cdot)$ is the indicator function.  
    \item \emph{Step 2.} Calculate the element-wise mean of the membership matrices $\bar{\mathcal{A}} = \frac{1}{L} \sum_{l=1}^{L} \mathcal{A}^{(l)}$.
    \item \emph{Step 3.} Identify the most \emph{representative} posterior draw as the one that is closest to $\bar{\mathcal{A}}$ with respect to the element-wise Euclidean distance $\sum_{i=1}^{n} \sum_{j=1}^{n} (\mathcal{A}^{(l)}(i,j) - \bar{\mathcal{A}}(i,j))^{2}$ among the retained $l = 1,\ldots,L$ posterior draws.
\end{itemize}

The posterior estimates of cluster memberships $Z_1,\ldots,Z_n$ and other model parameters $\bm{\Theta}$ can be also obtained using Dahl's method accordingly.

\section{Theory}\label{sec:theory}
Next we study the theoretical properties for the posterior distribution obtained from model \eqref{eq:MFM_MxN}. In order to establish the posterior contraction results, we consider a refined parameter space $\bm{\Theta^*}$ defined as
$\cup_{k=1}^{\infty} \bm{\Theta_k^*}$, where $\bm{\Theta_k^*}$ corresponds to the compact parameter space for all the model parameters (i.e., mixture weights, matrix normal mean and covariances) given a fixed cluster number $K=k$. More precisely, we define $\bm{\Theta_k^*}$ as
\begin{align*}
     \Big\{&w_1,\ldots,w_k \in (\epsilon, 1-\epsilon),~ \sum_{i}^k w_i = 1, ~~ M_1,\ldots,M_k \in (-C_1, C_1)^{p \times q}, \\
    & \sigma_{1}(U_i), \ldots, \sigma_p(U_i) \in (\underline{\sigma}, \bar{\sigma}), ~~ e_{1}(U_i),\ldots,e_p(U_i) \in (-C_2,C_2)^p ~~\text{for every}~i=1,\ldots,k,\\ & \sigma_{1}^*(V_j), \ldots, \sigma_q^*(V_j) \in (\underline{\sigma}, \bar{\sigma}), ~~ e_{1}^*(V_j),\ldots,e_q^*(V_j) \in (-C_3,C_3)^q ~~\text{for every}~j=1,\ldots,k.  \Big\},
\end{align*}
where $\epsilon, \underline{\sigma}, \bar{\sigma}, C_1,C_2,C_3$ are some positive constants,  $\{\sigma_1(U_i),\ldots,\sigma_p(U_i); e_1(U_i),\ldots,e_p(U_i)\}$, $\{\sigma_1^*(V_j),\ldots,\sigma_q(V_j); e_1^*(V_j),\ldots,e_q^*(V_j)\}$ are eigenvalues and eigenvectors for matrix $U_i$ and $V_j$, respectively. We also define the mixing measure as $G = \sum_{i=1}^k w_i \delta_{\gamma_i}$, where $\delta$ is the point mass measure, and $\gamma_i = \{M_i, U_i, V_i \}$ is the collection of parameters for the matrix normal distribution in cluster $i$ for $i=1,\ldots,k$. For two sequence of real numbers $\{a_n\}$ and $\{b_n\}$, we define $a_n \lesssim b_n$ if there exists a universal positive constant $C$ whose value is independent of $n$ such that $a_n \leq C b_n$. For any two mixing measures $G_1 = \sum_{i=1}^k p_i \delta_{\gamma_i}$ and $G_2 = \sum_{j=1}^{k'} p_j' \delta_{\gamma_j}$, we define their Wasserstein distance as $W(G_1,G_2) = \inf_{q \in \mathcal{Q}} \sum_{i,j} q_{ij} \|\gamma_i - \gamma_j\| $, where $\| \cdot \|$ is the element-wise $L_2$-distance, $\mathcal{Q}$ denotes the collection of joint discrete distribution on the space of $\{1,\ldots,k\} \times \{1,\ldots,k'\}$ and $q_{ij} $ is the probability being associated with $(i,j)$-element and it satisfies the constraint that $\sum_{i=1}^k q_{ij} = p_j'$ and $\sum_{j=1}^{k'} q_{ij} = p_i$, for every $i=1,\ldots,k$ and $j=1,\ldots,k'$. 

Let $K_0$, $G_0$, $P_0$ be the true number of clusters, the true mixing measure, and the corresponding probability measure, respectively. Then the following theorem establishes the posterior consistency and contraction rate for the cluster number $K$ and mixing measure $G$. The proof is given in Supplementary Materials, Section S6; and it is based on the general results for Bayesian mixture models in \citet{guha2019posterior}.  
\begin{theorem}\label{thm1}
Let $\Pi_n(\cdot \mid Y_1,\ldots,Y_n)$ be the posterior distribution obtained from \eqref{eq:MFM_MxN} given a random sample $Y_1,\ldots,Y_n$. Assume that the parameters of interest are restricted to $\bm{\Theta^*}$. Then we have \begin{align*}
\Pi_n(K = K_0 \mid Y_1,\ldots,Y_n) \rightarrow 1, ~\text{and}~~ \Pi_n (W(G,G_0)\lesssim (\log n/n)^{-1/2} \mid Y_1,\ldots,Y_n) \rightarrow 1, 
\end{align*}
almost surely under $P_0$ as $n \rightarrow \infty$. 
\end{theorem}
Theorem \ref{thm1} shows that our proposed Bayesian method is able to correctly identify the unknown number of clusters and the latent clustering structure with posterior probability tending to one as the sample size increases. The requirement of a compact parameter space $\bm{\Theta^*}$ is commonly used in the Bayesian nonparametrics literature \citep{guha2019posterior}, and it is practically relevant since the model parameters are expected to take values in a pre-specified range. For example, it is reasonable to assume that the mixture weights are greater than some extremely small number such as $.001\%$ to yield meaningful clustering results.

\section{Simulation}\label{sec:simu}
\subsection{Simulation Setup}\label{sec:setup}

We conduct simulation studies to examine the finite-sample performance of the proposed method based on three evaluation metrics, (i) probability of choosing the correct number of clusters, (ii) Rand index \citep{rand1971objective}, and (iii) root mean squared error in estimating $V \otimes U$. Those three metrics serve as useful evaluation measures in terms of the model selection accuracy, clustering structure recovery performance, and parameter estimation accuracy. 

We compare the performance of the proposed method with that of two classical benchmark methods, $K$-means clustering \citep{hartigan1979algorithm} and spectral clustering \citep{ng2002spectral}. Both methods take the vectorized matrices as the input. Those two benchmark methods are implemented using the built-in function \texttt{kmeans} and the function \texttt{specc} in R package \texttt{kernlab} \citep{kernlab} with a Gaussian kernel under default settings, respectively. The Rand index is calculated using the function \texttt{rand.index} in R package \texttt{fossil} \citep{vavrek2011fossil}. 

\begin{figure}
\centering
\def\tabularxcolumn#1{m{#1}}
\begin{tabularx}{\linewidth}{@{}cXX@{}}
\begin{tabular}{lcr}
     \subfloat[$M_1$]{\includegraphics[width=0.27\textwidth]{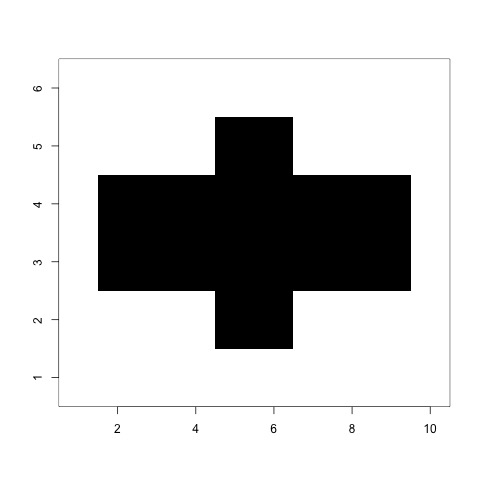}} 
   & \subfloat[$M_2$]{\includegraphics[width=0.27\textwidth]{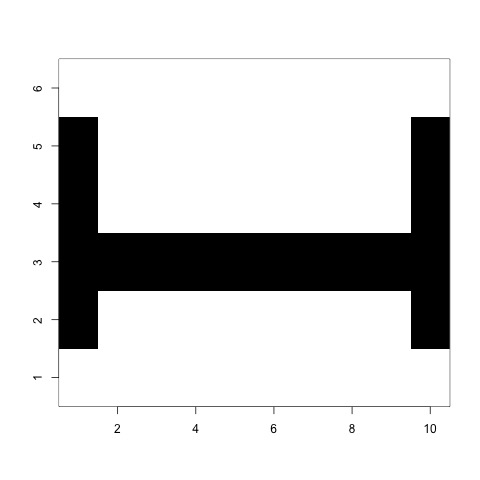}}
   & \subfloat[$M_3$]{\includegraphics[width=0.27\textwidth]{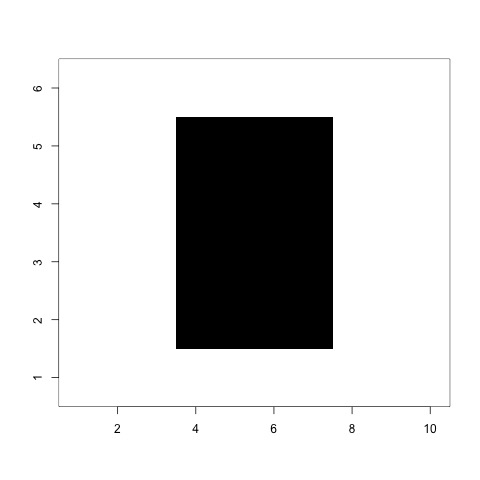}} \\
     \subfloat[$\hat{M}_1$, $n=100$]{\includegraphics[width=0.27\textwidth]{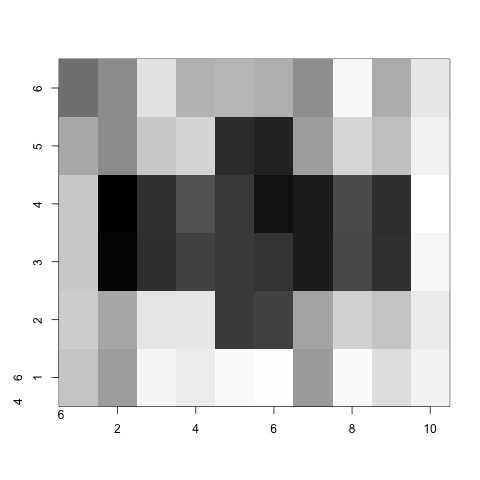}} 
   & \subfloat[$\hat{M}_2$, $n=100$]{\includegraphics[width=0.27\textwidth]{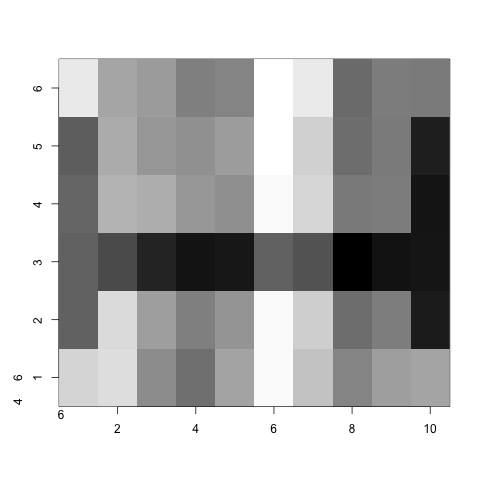}}
   & \subfloat[$\hat{M}_3$, $n=100$]{\includegraphics[width=0.27\textwidth]{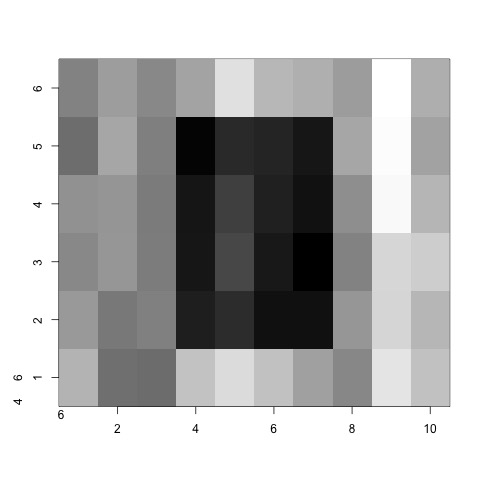}}\\
     \subfloat[$\hat{M}_1$, $n=200$]{\includegraphics[width=0.27\textwidth]{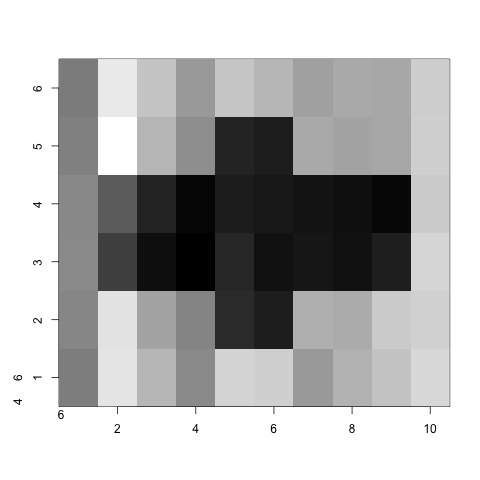}} 
   & \subfloat[$\hat{M}_2$, $n=200$]{\includegraphics[width=0.27\textwidth]{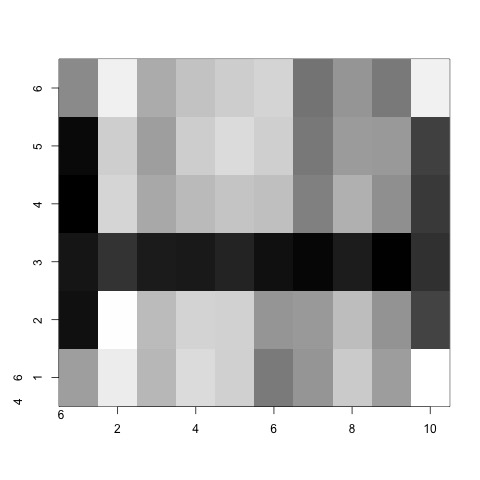}}
   & \subfloat[$\hat{M}_2$, $n=200$]{\includegraphics[width=0.27\textwidth]{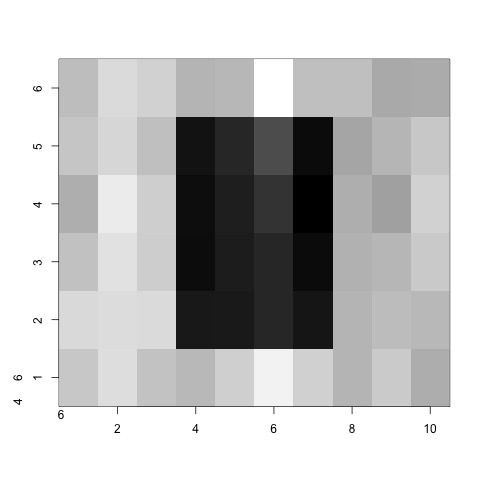}}\\
     \subfloat[$\hat{M}_1$, $n=400$]{\includegraphics[width=0.27\textwidth]{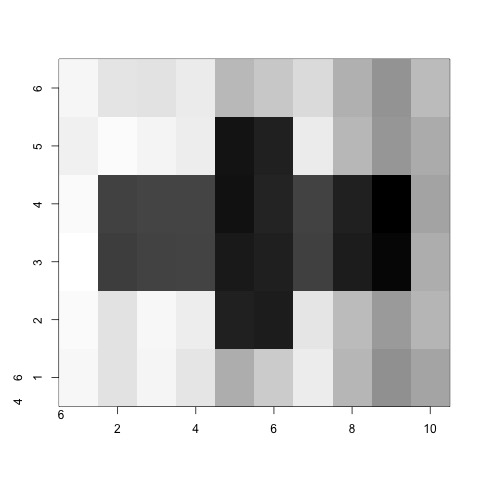}} 
   & \subfloat[$\hat{M}_2$, $n=400$]{\includegraphics[width=0.27\textwidth]{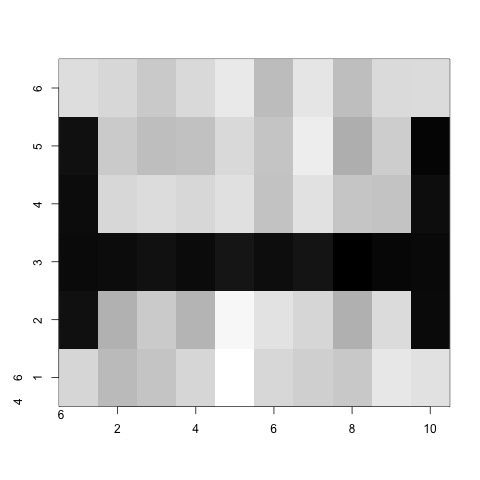}}
   & \subfloat[$\hat{M}_3$, $n=400$]{\includegraphics[width=0.27\textwidth]{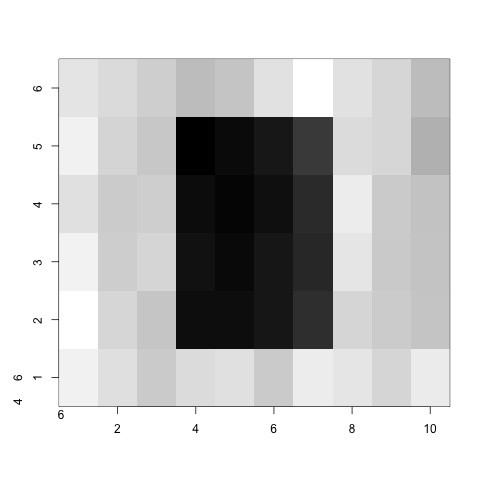}}\\
\end{tabular}
\end{tabularx}
\caption{True signals and representative draws from recovered signals by MFM under $n=100, 200, 400$ and high noise level ($V = \Sigma_{AR(1), 0.9, 6}$). }
\label{fig:shapes_10by6_simulation}
\end{figure}

When generating the data, we consider two matrix sizes: (i) small matrix size, where $p=10$ and $q=6$, and (ii) large matrix size, where $p=25$ and $q=18$. For small matrix size, we generate three clusters of signals from matrix normal distributions with weights $\pi = (0.3, 0.3, 0.4)$ and the mean matrices $M_1, M_2, M_3 \in \mathcal{R}^{10\times6}$ displayed in the first row of Figure \ref{fig:shapes_10by6_simulation}, where the elements are coded as $1$ if corresponding regions are shaded, and $0$ otherwise. The row-wise covariance matrix $U$ is drawn from a standard Wishart distribution with $\nu = 11$ and dimension $10$ (to ensure that the marginal variance of the noise is equal to $\sigma^2$, $U$ is converted to a correlation matrix), and the column-wise covariance matrix $V$ is a $6 \times 6$ AR(1) matrix with $\rho = 0.9$ (i.e., $V = \Sigma_{AR(1), 0.9, 6}$). We set the total sample size $n \in \{ 100, 200, 400\}$. To examine the performance of the proposed method and the other two competitive methods under different noise levels, we also consider another setting under which the row-wise covariance matrix $V = 0.5^2 \times \Sigma_{AR(1), 0.9, 6}$. We run $100$ Monte-Carlo replications, and for each replication we run MCMC chains for $1500$ iterations, where the first $1000$ draws are discarded as burn-in for the experiments on small size matrix. 

\begin{figure}[htp]
    \centering
    \includegraphics[width=0.8\textwidth]{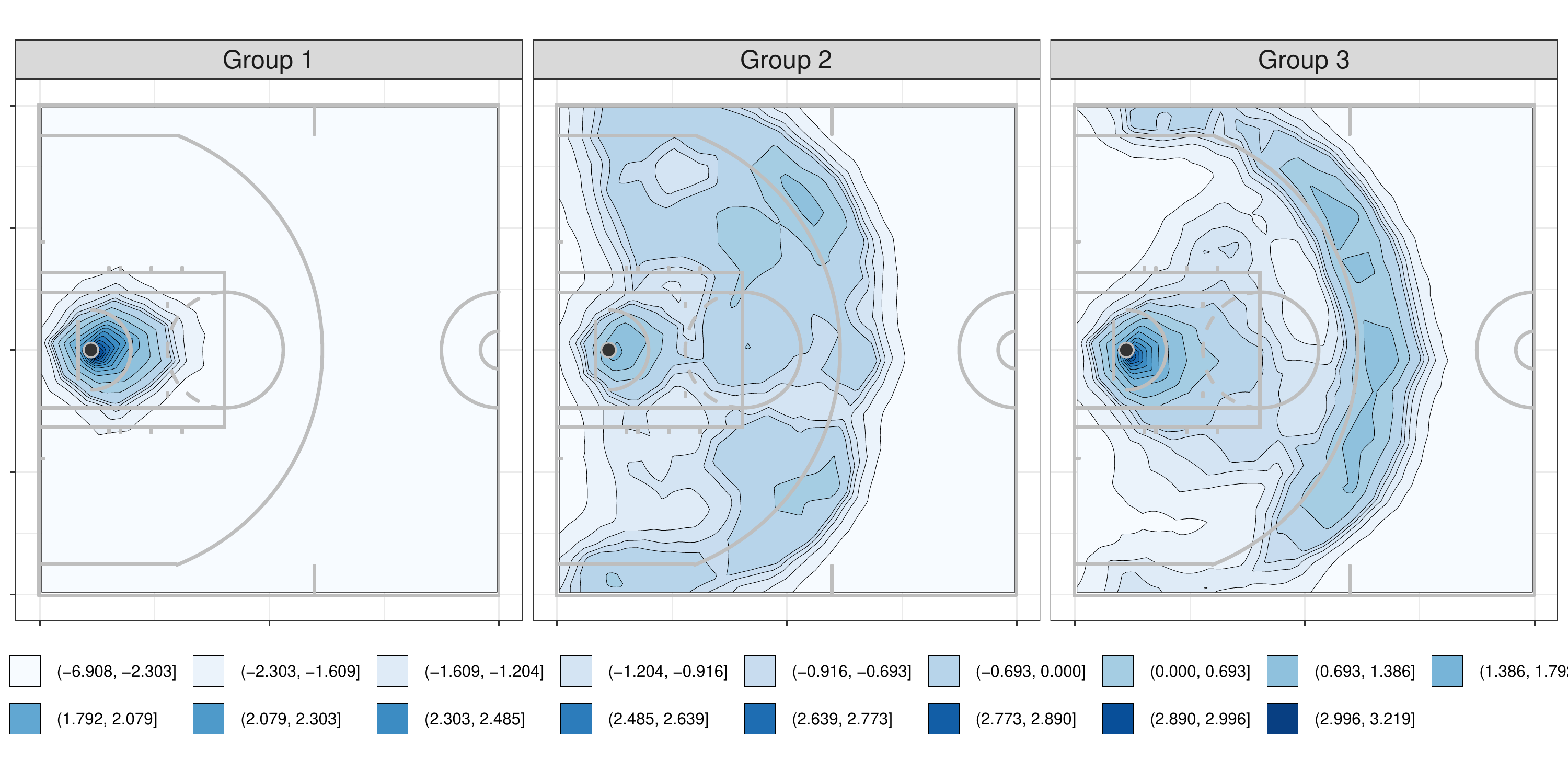}
    \caption{Log Intensity Maps of Three Patterns}
    \label{fig:log_inten_simu}
\end{figure}

For large matrix size, we consider three mean patterns shown in Figure \ref{fig:log_inten_simu}, each of which is designed to represent a prevalent offensive style in the NBA shooting chart data. From the log intensity maps in Figure~\ref{fig:log_inten_simu}, we note that the Group 2 represents all-around players such as LeBron James; Group 3 represents three-point shooters such as Eric Gordon; and Group 1 represents inside players such as Steven Adams. With the mean structure being the log intensity shown in Figure \ref{fig:log_inten_simu}, we generate our simulated data from matrix normal distributions with column-wise covariance matrix $V = \sigma^2 \times \Sigma_{AR(1), \rho, 25}$ and row-wise covariance matrix $U$ drawn from a standard Wishart distribution with $\nu = 19$ and dimension $18$ (to ensure that the marginal variance of the noise is equal to $\sigma^2$, $U$ is converted to a correlation matrix). We fix $n=200$ to mimic the number of players in the motivating data example, and choose $\pi=(0.3,0.4,0.3)$. We also run $100$ replicates and for each replicate we run a MCMC chain for $1200$ iterations, where the first $600$ draws are discarded as burn-in. 

To ensure a comparison that is as fair as possible, for each replication, the number of clusters for $K$-means and spectral clustering are chosen as the same number of clusters obtained from the proposed MxN-MFM. The MCMC settings are chosen based on pilot runs, and the overlaid traceplots of Rand index further justify the validity of the chosen MCMC settings, which are shown in the Supplementary Materials.
All computations presented in this paper were performed in R (version 3.6.0) \citep{R-Team} on a computing server (256GB RAM, with 8 AMD Opteron 6276 processors, operating at 2.3 GHz, with 8 processing cores in each). 

\subsection{Simulation Results}\label{sec:simu_results}
\subsubsection{Small matrix size results}

We first present the results for small matrix size ($p=10, q=6)$. Table \ref{tb:RI_10by6} shows the mean Rand index for three different methods under different sample size and noise levels. It is clear that the proposed method (MFM-MxN) outperforms $K$-means and spectral clustering under all scenarios, and its advantage is more salient when the noise level is higher, which is particularly important for clustering. The clustering accuracy (mean Rand index $> 0.95$) of the proposed MFM-MxN method is also compelling at the absolute scale, as the rule-of-thumb Rand index threshold value for ``good clustering'' is $0.80$. We then summarize the distribution of estimated number of clusters for MFM-MxN in Table \ref{tb:Khat_10by6}. We note that the probability of identifying the correct number of clusters is very satisfactory ($>80\%$) for the proposed MFM-MxN, and this probability increases as the sample size increases. These results confirm the benefit of taking account for the matrix structure and the flexibility of full Bayesian inference in the clustering analysis. 

\begin{figure}[htp]
\centering
\includegraphics[width=.8\textwidth]{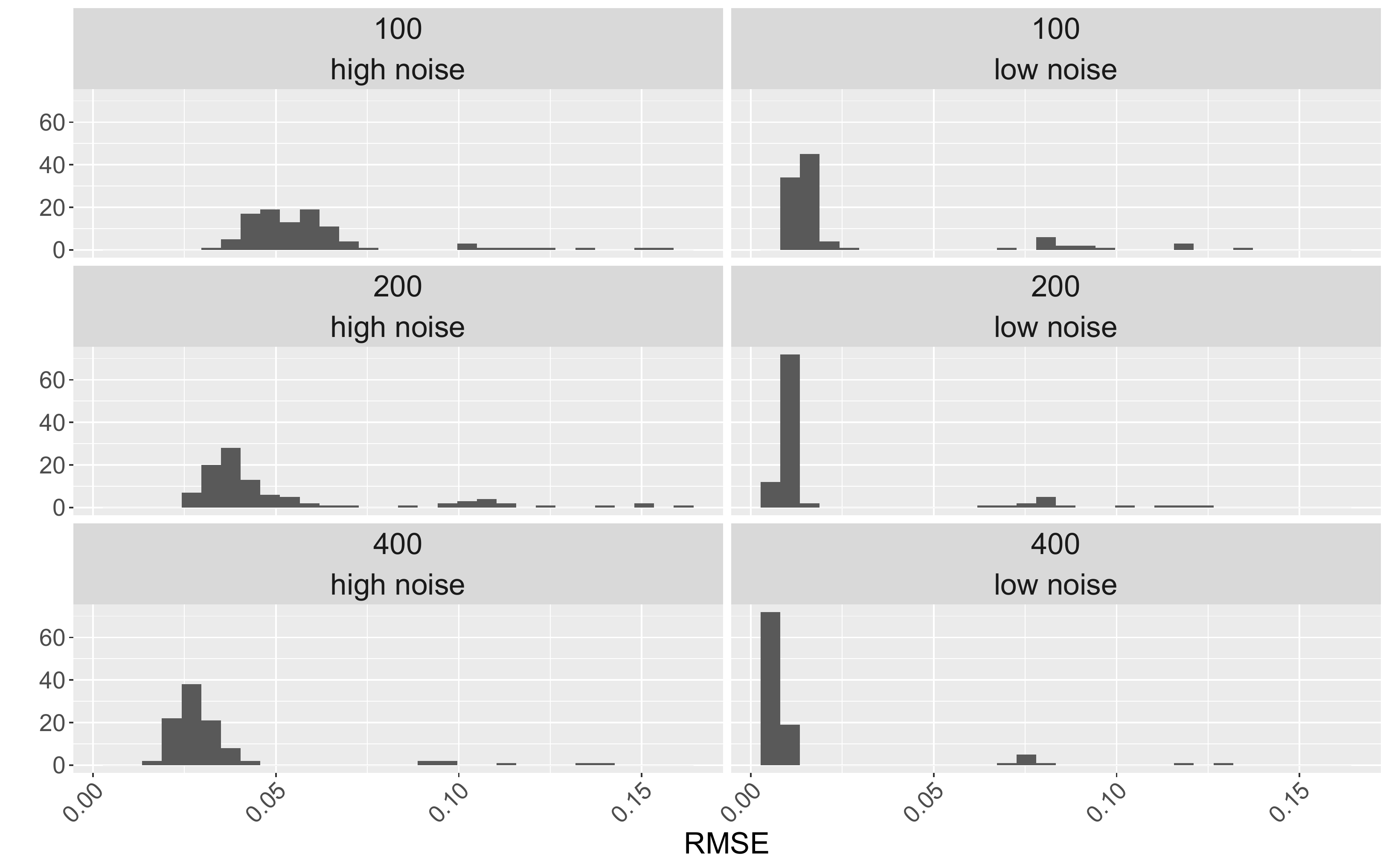}
\caption{Histograms of RMSE (over 100 replicates) for the covariance estimates ($\hat{V} \otimes \hat{U}$) under different sample sizes ($n=100, 200, 400$) and noise levels (high: $V = \Sigma_{AR(1), 0.9, 6}$, low: $V = 0.5^2 \times \Sigma_{AR(1), 0.9, 6}$). \label{fig:VU_RMSE_10by6}}
\end{figure}

We also present the RMSE for the estimation of covariance matrix $V \otimes U$ in Figure \ref{fig:VU_RMSE_10by6}. It is clear that the estimation accuracy improves as the noise level drops down and when the sample size increases. This is also confirmed in Figure \ref{fig:shapes_10by6_simulation} where the recovered signals are significantly less noisy and better recapitulate the true signals as the sample size increases. 


\begin{table}[htp]
\centering
\caption{Simulation results for small matrix size: Mean Rand index obtained from MFM-MxN, $K$-means and Spectral Clustering under different sample size and noise levels (high: $V = \Sigma_{AR(1), 0.9, 6}$, low: $V = 0.5^2 \times \Sigma_{AR(1), 0.9, 6}$) based on 100 Monte-Carlo replications. \label{tb:RI_10by6}}
\begin{tabular}{l|lll|lll}
 &  & $\sigma=1$ & & & $\sigma=0.5$ &  \\ 
  \hline
& MFM & $K$-means & Spectral & MFM & $K$-means & Spectral \\ 
  \hline
$n=100$ & $\bm{0.977}$ & 0.558 & 0.559 & $\bm{0.964}$ & 0.837 & 0.886 \\ 
$n=200$ & $\bm{0.958}$ & 0.550 & 0.552 & $\bm{0.967}$ & 0.846 & 0.911 \\ 
$n=400$ & $\bm{0.984}$ & 0.553 & 0.555 & $\bm{0.979}$ & 0.878 & 0.956\\ 
\bottomrule
\end{tabular}
\end{table}

\begin{table}[htp]
\centering
\caption{Simulation results for small matrix size: Percentage ($\%$) of selected number of clusters $\hat{K}$ for MFM-MxN under different sample sizes and noise levels (high: $V = \Sigma_{AR(1), 0.9, 6}$, low: $V = 0.5^2 \times \Sigma_{AR(1), 0.9, 6}$) based on $100$ Monte-Carlo replicates. The true number of clusters is 3. \label{tb:Khat_10by6}}
\begin{tabular}{l|lll|lll}
\toprule
 noise &  & high & & & low &  \\ 
\hline
$\hat{K}$& 2 & 3 & 4 & 2 & 3 & 4 \\ 
  \hline
$n=100$ &  10 &  $\bm{90}$ &   0 & 16 & $\bm{84}$ & 0 \\ 
$n=200$ &  18 &  $\bm{82}$ &   0 & 14 & $\bm{86}$ & 0\\ 
$n=400$ &   7 &  $\bm{93}$ &   0 & 9 & $\bm{91}$ & 0 \\ 
\bottomrule
\end{tabular}
\end{table}

\subsubsection{Large matrix size results}
In Tables \ref{tb:Khat_25by18} and \ref{tb:RI_25by18}, we present the results for large matrix size data ($p=25$, $q=18$). Similar to previous findings, our proposed method is able to correctly find the true number of clusters at least $80\%$ of the time for different settings. Our method also has the highest average Rand index ($> 0.95$), which is higher than that of the other two benchmark methods, indicating that the proposed method is very powerful in terms of recovering the latent clustering structure, even for matrix data of larger size.

\begin{table}[htp]
\centering
\caption{Simulation results for large matrix size: Percentage ($\%$) of number of clusters for MFM-MxN under different noise levels ($V = \sigma^2 \Sigma_{AR(1), \rho, 6}$) based on $100$ replicates. The true number of clusters is 3. \label{tb:Khat_25by18}}
\begin{threeparttable}
\begin{tabular}{l|lll|lll|lll}
& \multicolumn{3}{c|}{$\sigma=1.5$} & \multicolumn{3}{c|}{$\sigma=1.0$} & \multicolumn{3}{c}{$\sigma=0.5$}   \\ 
  \hline
& 2 & 3 & 4 & 2 & 3 & 4 & 2 & 3 & 4 \\ 
  \hline
$\rho=0.9$ &  11 & $\bm{89}$ & 0 & 11 & $\bm{89}$ & 0 & 11 & $\bm{89}$ & 0 \\ 
$\rho=0.6$ &  13 & $\bm{87}$ & 0 & 13 & $\bm{87}$ & 0 & 12 & $\bm{88}$ & 0 \\ 
$\rho=0.3$ &  0 & $\bm{94}$\tnote{\dag} & 2 & 13 & $\bm{87}$ & 0 & 11 & $\bm{89}$ & 0 \\ 
\bottomrule
\end{tabular}
\begin{tablenotes}
\centering
\footnotesize
\item[\dag] Results for four runs are not shown due to large estimated cluster numbers.
\end{tablenotes}
\end{threeparttable}
\end{table}

\begin{table}[htp]
\centering
\caption{Simulation results for large matrix size: Mean Rand index for MFM-MxN, $K$-means and Spectral Clustering under different noise level $\sigma$ and correlation strength $\rho$, based on 50 replicates. \label{tb:RI_25by18}}
\resizebox{\textwidth}{!}{\begin{tabular}{l|lll|lll|lll}
& & $\sigma=1.5$ &  & & $\sigma = 1.0$ & & & $\sigma = 0.5$ &   \\ \hline
 & MFM & $K$-means & Spectral & MFM & $K$-means & Spectral & MFM & $K$-means & Spectral  \\ 
  \hline
$\rho=0.9$ &   $\bm{0.963}$ & 0.917 & 0.918 & $\bm{0.963}$ & 0.910 & 0.934 & $\bm{0.959}$ & 0.905 & 0.951\\ 
$\rho=0.6$ &    $\bm{0.957}$ & 0.917 & 0.945 & $\bm{0.957}$ & 0.899 & 0.929 & $\bm{0.953}$ & 0.895 & 0.924 \\ 
$\rho=0.3$ &    $\bm{1.000}$ & 0.959 & 0.984  & $\bm{0.957}$ & 0.917 & 0.946 & $\bm{0.953}$ & 0.923 & 0.930 \\ 
\bottomrule
\end{tabular}}
\end{table}



\section{Application to NBA Shot Chart Data analysis}\label{sec:real_data}
In this section, we apply the proposed method to investigate the shooting pattern of players in the 2017-2018 NBA regular season. 
Our analysis is purely based on the location of shots and hence the resulting clusters are completely data-driven without considering other information about players or their teams. The shots that are made 36 ft away from the baseline are not included in this analysis as they are usually not part of the regular tactics. We start by obtaining the intensity surface\footnote{The resulting intensity surface is scaled to adjust for the number of games played by the respective player.} for each player by fitting an LGCP to raw shot location data using off-the-shelf functions in R package \texttt{inlabru} \citep{inlabru}. The logarithm of each intensity surface is then discretized to generate a $25$ by $18$ matrix as the main variable of interest.

To implement our method, we run $50$ independent MCMC chains with random initial values, each of which has $6000$ iterations where the first $4000$ are discarded as burn-in to ensure the convergence. We select a representative chain that has the highest mean concordance value (mean Rand index = $0.82$) compared to all other chains with respect to the clustering memberships. This representative chain yields 3 groups of size 71, 23 and 97, respectively. Visualizations of intensity matrices with contour for selected players from these three groups are presented in Figure~\ref{fig:real_data_group}. A full list of  player names for each group are given in the Supplementary Materials, Section S1. We also plot the estimated covariance matrices in Section S7. We find that both the column-wise covariance $\hat{U}$ and row-wise covariance matrix $\hat{V}$ enjoy a banded structure, i.e., the correlations excluding the diagonal and off-diagonal entries are quite small, which confirms that our method is able to take spatial/location information into account by modeling the matrix structure in the data appropriately. 
\begin{figure}[htp]
    \centering
    \includegraphics[width=0.7\textwidth]{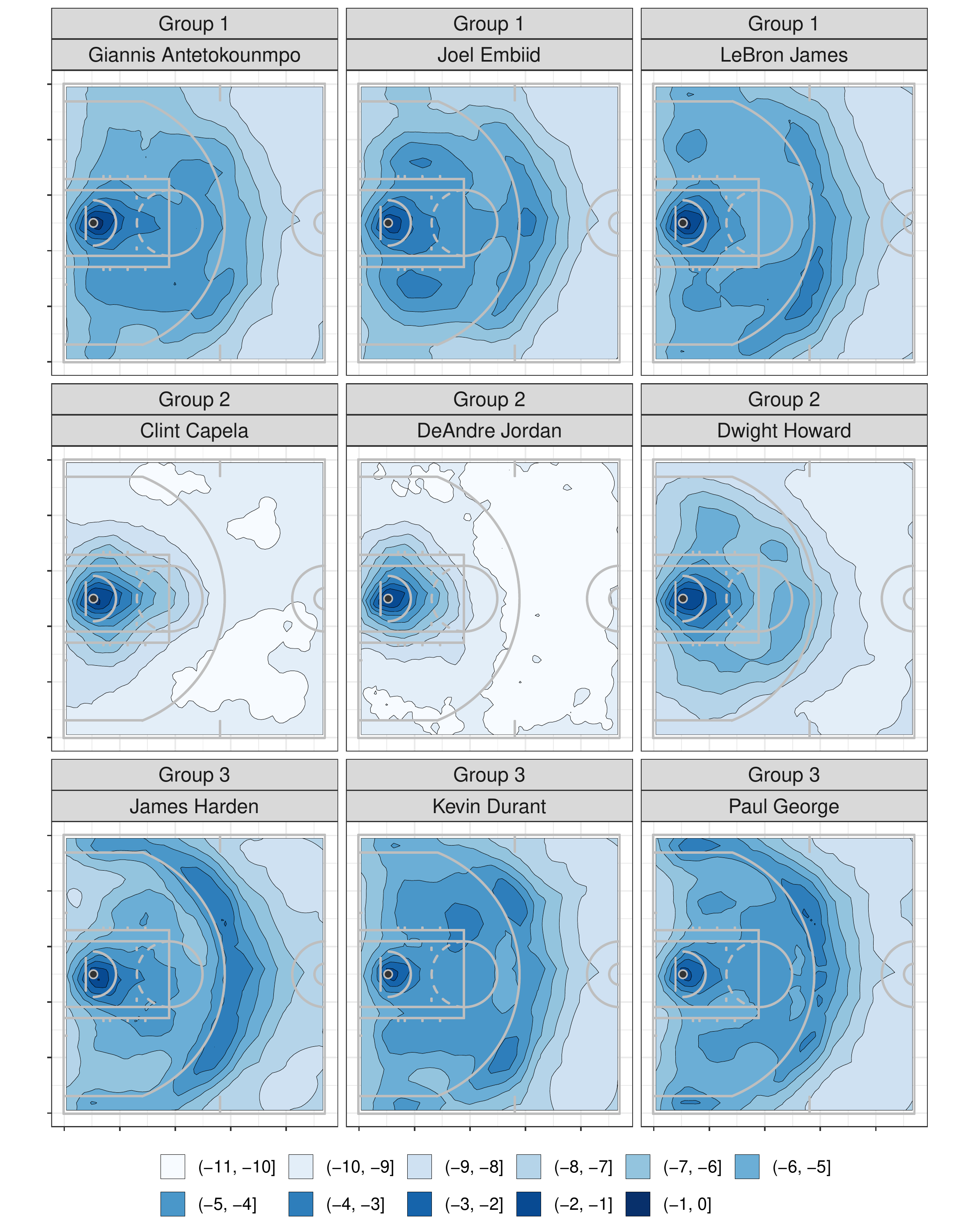}
    \caption{Log Intensity Matrices with Contour for Selected Players}
    \label{fig:real_data_group}
\end{figure}

Several interesting observations can be made from the visualization results. We see that for the players in Group 1, they are able to make all types of shots, including three-pointers, perimeter shot, and also shots over the painted area. However, compared with the players in Group 3, they have less three-pointers. Most players in this group are Power Forward and Small Forward. However, we can still find some players such as Dwyane Wade, Ricky Rubio and Rajon Rondo in this group. This can be explained by the fact that the three players mentioned above do not have a good three-point shot percentage and tend to attack the basket in paint area. The players in Group 2 have most shots located near the hoop. They are good at making alley-oops and slam dunks, however, not good at making three-pointers. Most of them are Center. There is still an interesting player in this group, Dejounte Murray. He plays very similarly with Tony Park who is a previous player of Spurs and makes more field goal attempts on paint area like the Center. For the players in Group 3, we find that they have more three-pointers compared to the other two groups, and the players in this group are almost all Shooting Guard. Although we still find that Kevin Durant belongs to this group, because he is an all-rounder and has an excellent ability of scoring three-pointers. In addition, there are some inside players in this group such as Kevin Love, Kelly Olynyk, and DeMarcus Cousins. This reflects the recent trend that the NBA teams start to prefer or even require their inside players to shoot from downtown and release the space in paint area. Our findings also confirm that the number of pure inside players decreases as the three-pointer becomes a conventional weapon for most players. 
\section{Discussion}
\label{sec:discussion}
In this paper, we propose a novel Bayesian nonparametric clustering approach for learning the latent heterogeneity in matrix-valued response data. Building upon the mixture of finite mixtures framework, we develop a collapsed Gibbs sampler for efficient Bayesian inference and adopt Dahl's method for post MCMC inference. Numerical results have confirmed that the proposed method is able to simultaneously infer the number of clusters and the model parameters with high accuracy. 
Comparing to the traditional clustering techniques such as $K$-means and spectral clustering, the proposed method is able to improve the clustering performance especially when the noise level is high for the reason that the rich spatial location information is incorporated by handling the data in the matrix format. 

In the analysis of the NBA shot charts data, three prevalent shooting patterns along with the respective players are identified. The results provide valuable insights to both players and managers - players can obtain more comprehensive understandings of their current attacking patterns, and hence develop further training plans accordingly; the managers can be equipped with more \emph{objective} and principled analysis of shooting patterns of the players in the league, and hence make better data-informed decisions on player recruiting. 


A few topics beyond the scope of this paper are worth further investigation. A natural extension is Bayesian clustering for general multi-way data such as tensors. Also, as matrix inverse is required at each MCMC update, the proposed estimation procedure can be slow when the matrix size is very large. Proposing an efficient algorithm for large matrix data devotes an interesting future work, and we envision low-rank approximation and sparse compression as two promising directions to mitigate this computational challenge. Finally, jointly estimating intensity surface and grouping information is another interesting direction for future work.   

\section*{Acknowledgements} 
The authors would like to thank Dr.~Yishu Xue and Dr.~Hou-Cheng Yang for providing the organized data which include the raw shot charts and estimated intensity maps via INLA and R code for data visualization.

\section*{Supplementary Materials}\vspace{-2mm}
Technical details about the posterior derivation, proof of theorems, additional numerical results are provided in the Online Supplementary Materials. R code and the data for the computations of this work are available at https://github.com/fyin-stats/MFM-MxN.

\bibliographystyle{abbrvnat}
\bibliography{sample.bib}

\end{document}